\documentclass[10pt]{iopart}
\usepackage[]{cite}
\usepackage[pdftex]{graphicx}

\usepackage[T1]{fontenc} 
\usepackage{bm}
\usepackage{array}
\usepackage[caption=false,font=footnotesize,labelfont=sf,textfont=sf]{subfig}
\usepackage{url}
\usepackage{hyperref}

\makeatletter
\def\bibinfo#1{%
  \@ifundefined{bibinfo@X@#1}%
    {\@firstofone}
    {\csname bibinfo@X@#1\endcsname}}
\makeatother

\newcommand*{\mytitle}{Control of sensitivity in vortex-type magnetic tunnel junction magnetometer sensors by the pinned layer geometry}
\newcommand*{\myshorttitle}{Control of sensitivity in vortex MTJ sensors}

\begin{document}
\title[\myshorttitle]{\mytitle}

\author{Motoki~Endo$^{1}$, Muftah~Al-Mahdawi$^{2,3}$, 
Mikihiko~Oogane$^{1,2,3}$, and Yasuo~Ando$^{1,2,3}$}

\address{$^1$ Department of Applied Physics, Tohoku University, Sendai 980-8579, Japan}
\address{$^2$ Center for Science and Innovation in Spintronics (Core Research Cluster), Tohoku University, Sendai 980-8577, Japan}
\address{$^3$ Center for Spintronics Research Network, Tohoku University, Sendai 980-8577, Japan}
\ead{mahdawi@tohoku.ac.jp}
\vspace{10pt}
\begin{indented}
\item[]Corresponding authors: M.~Endo (e-mail: endo@mlab.apph.tohoku.ac.jp), M.~Al-Mahdawi (e-mail: mahdawi@tohoku.ac.jp).
\item[]January 2022
\end{indented}
\begin{abstract}
   The tuning of sensitivity and dynamic range in linear magnetic sensors is required in various applications. We demonstrate the control and design of the sensitivity in magnetic tunnel junction (MTJ) sensors with a vortex-type sensing layer. In this work, we develop sensor MTJs with NiFe sensing layers having a vortex magnetic configuration. We demonstrate that by varying the pinned layer size, the sensitivity to magnetic field is tuned linearly. We obtain a high magnetoresistance ratio of 140 \%, and we demonstrate a controllable sensitivity from 0.85 to 4.43 \%/Oe, while keeping the vortex layer fixed in size. We compare our experimental results with micromagnetic simulations. We find that the linear displacement of vortex core by an applied field makes the design of vortex sensors simple. The control of the pinned layer geometry is an effective method to increase the sensitivity, without affecting the vortex state of the sensing layer. Furthermore, we propose that the location of the pinned layer can be used to realize more sensing functionalities from a single sensor.
\end{abstract}

\vspace{2pc}
\submitto{\JPD}
\maketitle
\ioptwocol

\section{Introduction}
    Semiconductor Hall sensors are the mainstream technology for linear magnetometer sensors \cite{review_hirohata_2020}. Magnetoresistive (MR) sensing technologies, such as anisotropic MR, giant MR, and tunneling MR (TMR) are finding an increased use, due to the evolving requirements in automotive, internet-of-things (IoT), and biomedical applications \cite{review_zheng_2019}. Various requirements of high sensitivity, wide dynamic range, low nonlinearity, low power consumption, and low noise are required for covering the needs of the emerging applications.
    As an example of power sensing, an IoT device needs to monitor its battery current in the range of $<$100 nA in sleep mode, to $1.0$ A in communication mode \cite{hertlein_2018,TI_CC2650MODA}. Some devices may need a granular monitoring of critical subsystems such as the power harvester, the RF transceiver, \emph{etc}. Such applications require low-power miniature sensors that can measure in over 7--8 orders of magnitude. Automotive applications also have multiple sensitivity requirements for angle, position, and power sensing. Therefore, there is a need for a linear magnetometer sensor that can be designed into various dynamic ranges.
    
    The sensitivity in single MR elements, is usually designed by the material choice of the ferromangetic sensing layer, which set the magnetic anisotropy field \cite{fujiwara_2012_13,nakano_2017}. {High-sensitivity sensors are made from low-anisotropy soft ferromagnetic layers, which form domains and domain walls. The creep and depinning avalanches of domain walls are a significant source of auto-correlated noise \cite{ledoussal_2009,ferrero_2017}, and coercivity increase.}
    For obtaining a geometrically-designed sensitivity, a suitable magnetic texture is the magnetic vortex \cite{behncke_2018}. Magnetic vortices are found in various magnetic micro-structures, such as cross-tie domain walls \cite{metlov_2001,wiese_2007,mccord_2009}, confined domain walls \cite{meier_2007a}, and most importantly in micro-sized circular disks \cite{shinjo_2000}, and rectangles \cite{vogel_2011,miyake_2013}. They are characterized as flux-closure curling magnetization patterns with an out-of-plane magnetization at the core position, and they form due to a minimization of magnetization surface charges against the exchange energy. 
    Vortex-type magnetic sensors have been in interest recently \cite{novosad_2010, raberg_2018}. {Additional merits of using a vortex configuration is the possibility of lower noise due to topological protection \cite{suess_2018, he_2020}, and hysteresis is vanishingly small \cite{ostman_2014,he_2020}.}
    
    The vortex sensitivity is determined by the saturation field $H_\mathrm{an}$, at which the vortex core is at the edge of the disk. The ratio between free layer thickness and radius $(\beta = L_b/R_b)$ is the geometric parameter for setting $H_\mathrm{an}$ \cite{guslienko_2001a}. For the devices we will consider, $\beta \ll 1$ and $H_\mathrm{an}$ is linearly proportional to $\beta$ \cite{guslienko_2002}. By increasing the sensor free layer diameter, the sensitivity can be increased. However, above a critical diameter usually near 10 $\mu$m for soft magnetic layers \cite{he_2020}, the vortex state will break into a multi-domain state. The $H_\mathrm{an}$ is then limited to a minimum bound of 80--100 Oe.
    The TMR effect in MTJ sensors results in a large resistance readout sensitivity. The combination with a soft ferromagnetic free layer is the best candidate for fabricating vortex-type sensors.
    However, for a typical TMR ratio of 160\%, the maximum sensitivity (TMR$/2H_\mathrm{an}$) for vortex MTJ sensors is capped at 1 \%/Oe. Therefore, another control parameter for vortex-type magnetic sensors is needed. In this work, we show a method for sensitivity increase by constraining the pinned layer diameter. Hence, the sensitivity can be designed separately from the design of the vortex free layer, while keeping the linearity and low hysteresis.

\section{Fabrication and Methods}
    
    \begin{figure}
        \includegraphics[width=0.5\textwidth]{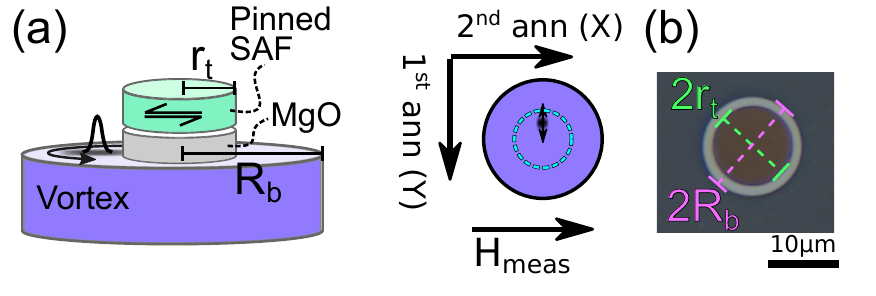}
        \caption{(a) A schematic of the sensor geometry and coordinates definitions. The core position sweeps linearly through the area covered by the pinned layer, marked by a dashed line.  (b) A microscope photograph of a representative element, during microfabrication.}
        \label{fig:schem}
    \end{figure}
    
    We fabricated vortex MTJ sensor devices as depicted in Fig.~\ref{fig:schem}(a).
    The vortex is formed in a circular free layer of a diameter {$2R_b = 10\mu$m}. The pinned layer for detection is at top with a smaller diameter of {$2 r_t = 2.0$--$9.5\mu$m}. As the vortex core moves linearly in response to magnetic field, the average inplane magnetization in the area enclosed by $2 r_t$ is also linearly dependent on the applied field. Therefore, we can decrease the dynamic range and increase sensitivity by reducing the area covered by $2 r_t$. The free layer requires a soft magnetic property, with a small magnetic anisotropy and small exchange stiffness, and we choose permalloy (NiFe) for this purpose. The magnetization in the pinned layer should be parallel to the sensing axis, but with a minimal stray field that does not affect the vortex response to measured field.
    We used radio-frequency magnetron sputtering to deposit MTJ films similar in design to previous reports \cite{fujiwara_2012_13}: thermally-oxidized silicon substrate/Ta 5/Ru 10/Ta 5/Ni$_{80}$Fe$_{20}$ 100/Ru 0.85/Co$_{40}$Fe$_{40}$B$_{20}$ 3/MgO $2.0$/Co$_{40}$Fe$_{40}$B$_{20}$ 3/Ru 0.85/Co$_{75}$Fe$_{25}$ 3.4/IrMn 10/Ta 3/Ru 5, where the numbers are the nominal thicknesses in nanometers. We optimized the top synthetic-antiferromagnet (SAF) pinned layer for a vanishing stray mangetic field \cite{NoteSupp}. The magnitude of TMR ratio is 140 \%, and the blanket films magnetization curves are typical for such a stack \cite{NoteSupp,fujiwara_2012_13}. During deposition, a static magnetic field was applied in the inplane direction using a permanent magnet on the holder stage {(defined as $Y$ direction)}, to set the induced magnetic anisotropy axis of NiFe layer \cite{NoteSupp, katada_2000,chikazumi_1955}. We micro-fabricated MTJ vortex sensor elements, using electron-beam (EB) lithography and Ar ion etching. During EB lithography exposures, we used in-field gold alignment markers for patterning alignment, {achieving a misalignment error less than 30 nm}. First, we fabricated and etched the bottom disk with a fixed diameter $2R_b = 10$ $\mu\mathrm{m}$.
    Then, we used a second EB lithography step to define the top disk with a varying diameter {$2 r_t = 2.0, 4.0, 5.0, 6.0, 8.0, 9.5$ $\mu\mathrm{m}$}, and etched the top disk down to the MgO layer.
    We monitored the etching processes with \emph{in situ} secondary-ion mass spectrometry, to monitor etching rate.
    The etching was at 30$^\circ$ off-normal incidence, for a sharp side profile. After that, we insulated the pillars with SiO$_2$, and deposited contact electrodes. A microscope image of a representative device during a different fabrication run is shown in Fig.~\ref{fig:schem}(b). After microfabrication, we applied a two-step magnetic annealing process at a 10-kOe magnetic field for 1 hour, with directions shown in Fig.~\ref{fig:schem}(a) \cite{fujiwara_2012_13}. The first annealing was at $350^\circ\mathrm{C}$ to crystallize CoFeB and obtain high TMR ratio. { The pinning field during the first annealing was at the same direction as the field applied during deposition ($Y$ direction) \cite{NoteSupp}}. The second pin annealing was at {a lower temperature $300^\circ\mathrm{C}$ to rotate the pinned layer direction $90^\circ$ and define the sensing axis ($X$ direction)}. This two-step process improves response linearity, while keeping the NiFe induced anisotropy \cite{fujiwara_2012_13}, which stabilize vortex nucleation along domain walls of reversal domains \cite{NoteSupp}.
    
    We simulated the static magnetization curves using OOMMF micromagnetic simulator \cite{oommf}. We simulated a circular disk with a diameter of $2R_b = 800$ nm and a thickness of 80 nm. We calculated the average magnetization within a smaller diameter $2r_t$, to be equivalent to the TMR measurements using the smaller pinned top layer. We used the following parameters, which represent soft magnetic properties: a saturation magnetization of $M_s = 800$ emu/cm$^3$, an exchange stiffness constant of $A_\mathrm{ex} = 13$ pJ/m, {zero crytalline magnetic anisotropy}, and a discretization cell of $2\times 2 \times 40$ nm$^3$. 
    The magneotostatic field of vortex state is from the side-surface magnetic charges, which is the same as a simple magnetic dipole \cite{guslienko_2001a,vogel_2010}. An inplane applied field $H_x$ will result in a linear orthogonal movement of the vortex core and a linear change of the pillar average magnetization. The saturation field $H_\mathrm{an}$ is inversely proportional to the zero-field susceptibility $\chi_0$. The static magnetic properties depend on the normalized $H_x/H_\mathrm{an}$, regardless of the pillar diameter. Therefore, a quantitative comparison between our simulations and experiments can be made after normalization.
    
\section{Results and Discussions}
    
    \begin{figure}
    \includegraphics[width=0.5\textwidth]{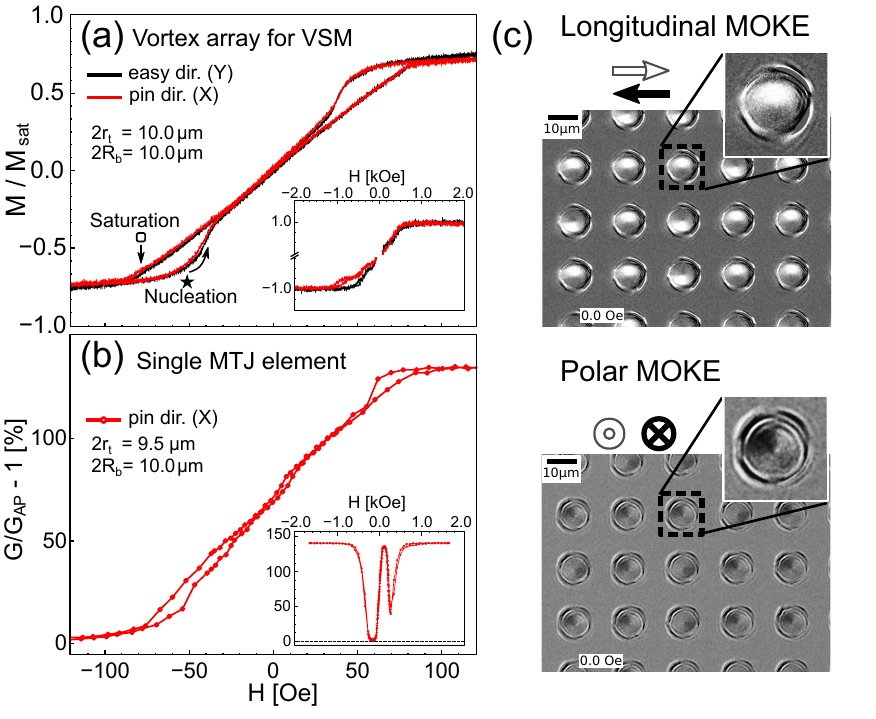}
    \caption{(a) Magnetization-field loops of 10-$\mu$m circular disks arranged in an array. (b) The conductance-field loop of a single-disk MTJ. (c) Longitudinal and polar MOKE images {at zero field after etching until the} NiFe free layer of the same sample in (a). The white and black arrows indicate the sense of color.}
    \label{fig:vsm_mr}
    \end{figure}

    In Fig.~\ref{fig:vsm_mr}, we show the verification of the vortex formation. First, we microfabricated {an array of $400 \times 400$ disks spanning $8 \times 8$ mm$^2$ area, where each disk is 10 $\mu$m in diameter, and the pitch is $40\mu$m.} We measured the $M$-$H$ loops by a vibrating-sample magnetometer (VSM). The center-to-center disks separation is four times the radius, and the inter-dot coupling can be neglected \cite{mejia-lopez_2006,vogel_2010}. {Therefore, the measured magnetization is of an average independent disk. In fig.~\ref{fig:vsm_mr}(a),} the $M$-$H$ curves show the typical hysteresis loops of vortex nucleation and linear displacement \cite{guslienko_2002}. The $M$-$H$ loops measured along the directions of the first annealing ($Y$) and second annealing ($X$) have the same loop shape and slope near zero field. The circular symmetry makes the magnetization hysteresis loops independent from $H$ direction.
    {The smooth transition at vortex nucleation shoulder indicates that the nucleation field has a distribution between 40--60 Oe, due to variation in edge roughness. The saturation field is tightly distributed at $H_\mathrm{an}$ = 79 Oe, since it is determined by the magnetostatic energy without effect from edge roughness.}
    In Fig.\ref{fig:vsm_mr}(b), we show the dependence of the tunneling conductance ratio $\left(G(H)/G_\mathrm{AP} - 1 \right)$ on applied field, using a single MTJ device prepared by the EB lithography process. The tunneling conductance is directly proportional to the magnetization average relative angle \cite{slonczewski_1989,nakano_2017,nakano_2018,ogasawara_2019a}. Both of the $G$-$H$ and $M$-$H$ loops match in shape, and the saturation fields are nearly equal at 83 Oe for the $G$-$H$ loop, and 79 Oe for the $M$-$H$ loop.
    The reorientation fields of the SAF pinned and free layers are at much higher fields [Fig.~\ref{fig:vsm_mr}(a,b) insets], and do not affect the vortex magnetization process. Due to the large thickness of NiFe compared to CoFeB layer, the Zeeman, exchange, and magnetostatic energies are dominated by the NiFe layer. Thus, the CoFeB free layer has anti-parallel vortex configuration to that of the NiFe layer. 
    For the same array of Fig.~\ref{fig:vsm_mr}(a), we etched the sample to expose the NiFe surface. We imaged the vortex magnetic configuration by a magneto-optical Kerr effect (MOKE) microscope [Fig.~\ref{fig:vsm_mr}(c)]. Before MOKE measurement, we applied a 10-kOe field along $X$ and out-of-plane directions in an external electromagnet. The domain images were taken as the difference from a reference image taken at 50 Oe. The longitudinal MOKE image shows the vortex curling inplane magnetization configuration. The polar MOKE image shows an out-of-plane magnetization at the center of the disks, indicating a perpendicularly-polarized vortex core. {The core polarization flips together with the vortex chirality. The fixing of polarity-chirality product indicates the presence of a bulk-type Dzyaloshinskii-Moriya Interaction \cite{im_2012}, which requires an extended investigation.} The presence of a small induced anisotropy affects the initial curling in vortex nucleation, and we find that the vortex formation is stable up to $2R_b = 30 \mu$m \cite{NoteSupp}.

    \begin{figure}
        \includegraphics[width=0.5\textwidth]{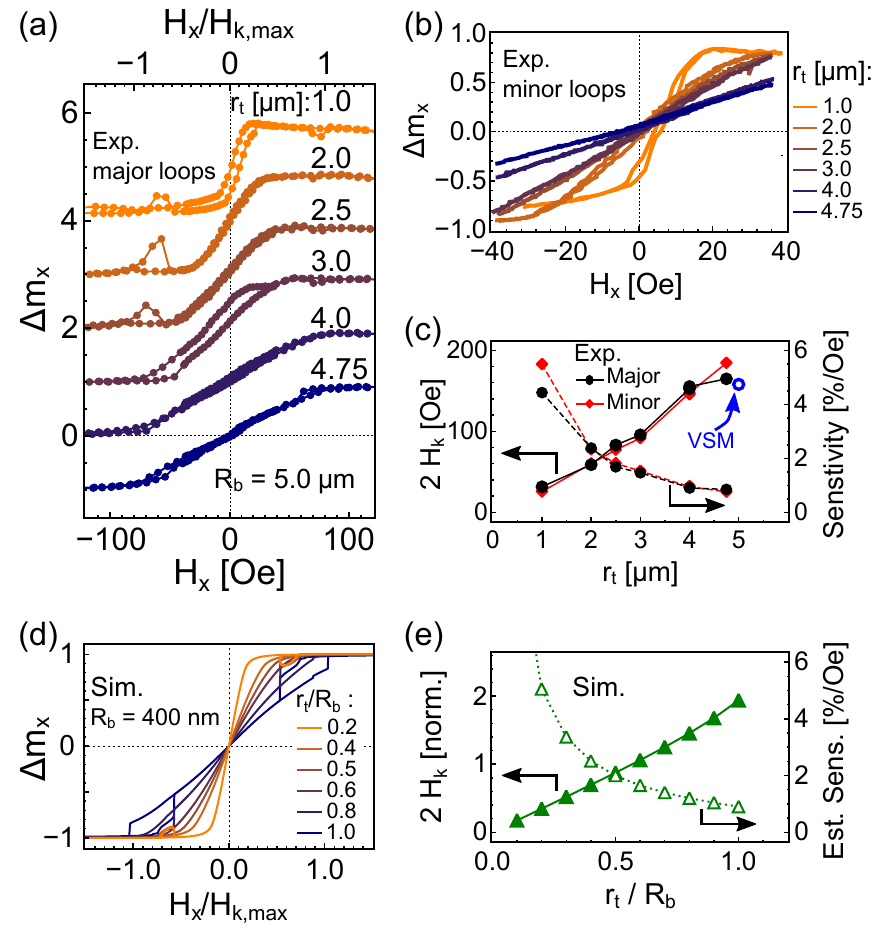}
        \caption{The effect of top disk size on sensitivity. (a) The experimental normalized magnetization $\Delta m_x$ major loops. The curves are vertically-shifted for clarity. (b) The corresponding minor loops. (c) The dependence of effective dynamic range $2 H_k$ and sensitivity on $r_t$. (d) The simulated $\Delta m_x$ loops. (e) The simulated dyanmic range and empirically-estimated sensitivity. }
        \label{fig:rt}
    \end{figure}
    
    Figure \ref{fig:rt} shows the control of sensitivity by changing the top disk radius $r_t$, which is the main result of this work. From the normalized conductance, we calculate the average normalized magnetization projected to the pinned-layer direction ($\Delta m_x$) in the area covered by the pinned layer disk, as follows:
    \begin{equation}
        \Delta m_x = 2 \, \frac{G - G_\mathrm{AP}}{G_\mathrm{P} - G_\mathrm{AP}} - 1,
    \end{equation}
    where $G_\mathrm{AP}$ and $G_\mathrm{P}$ are the conductances in the anti-parallel and parallel saturated configurations, respectively. In Fig.~\ref{fig:rt}(a), we show the experimental major loops. As we reduce $r_t$, the $\Delta m_x$-$H_x$ slope increases. The linearity of $\Delta m_x$-$H_x$ is preserved in the minor loops [Fig.~\ref{fig:rt}(b)]. In Fig.~\ref{fig:rt}(c), we show the dependence of the effective dynamic range $2H_k$ on $r_t$, estimated from linear fittings to the major and minor loops.
    The smallest $2H_k$ is $31$ Oe at $r_t = 1.0 \mu$m, which is in agreement of a linear scaling from $2H_{k, \mathrm{max}} = 156$ Oe obtained at $r_t = R_b = 5.0$ $\mu$m. Correspondingly, the sensitivity defined as $\mathrm{TMR}/2H_k$ increases 5 times from 0.85 $\%/\mathrm{Oe}$ to 4.43 $\%/\mathrm{Oe}$.
    
    In the micromagnetic simulations, the dynamic range is $2 H_{k, \mathrm{max}} = 1800$ Oe, due to the smaller $R_b = 400$ nm. However, after normalizing $H_x$ by $H_{k, \mathrm{max}}$, the simulated $\Delta m_x$-$H$ curves show the same properties and the increase of slope by decreasing $r_t$ as the experimental curves [Figs.~\ref{fig:rt}(d)]. For a quantitative comparison, we deduce the simulated sensitivity using the empirical values of TMR ratio of 140 \% and $2H_{k, \mathrm{max}} = 156$ Oe. We find the simulated sensitivity to be 0.90$\%/\mathrm{Oe}$ at $r_t/R_b = 1.0$, and increases to 5.0$\%/\mathrm{Oe}$ at at $r_t/R_b = 0.2$ [Fig.~\ref{fig:rt}(e)], in agreement with experimental results. The decrease in $2 H_K$ follows a simple linear relation with $r_t$, namely $H_k / H_{k, \mathrm{max}} = r_t / R_b$, both in simulations and experiments. This linear scaling relation is due to the linear displacement of vortex core by the applied field, until the core is located at $r_t$ when $H_x = H_k$. More improvement in sensitivity should be achievable by further reductions in $r_t$.
    {However, as $r_t$ is reduced, the stray field from the pinned SAF at the center of the disk increases, even when SAF magnetization is fully compensated at the blanket film level \cite{devolder_2019}. This can be seen as a field offset in the $\Delta m_x$-$H_x$ curves, which starts to appear at $r_t = 2.0$ $\mu$m, and become more pronounced at $r_t = 1.0$ $\mu$m. Further optimizations of the pinned SAF stack and etching process are required to minimize the effect of stray field. At $r_t/R_b \leq 0.6$, there are small dips near $H_x = \pm50$ Oe in Fig.~\ref{fig:rt}(a), and at $H_x/H_{k,\mathrm{max}} = 0.6$ in Fig.~\ref{fig:rt}(d). These are related to the curling in pre-nucleation state \cite{NoteSupp}. However, they should not affect the sensor performance, as they are outside the dynamic range of the sensor.}
    
    \begin{figure}
        \includegraphics[width=0.5\textwidth]{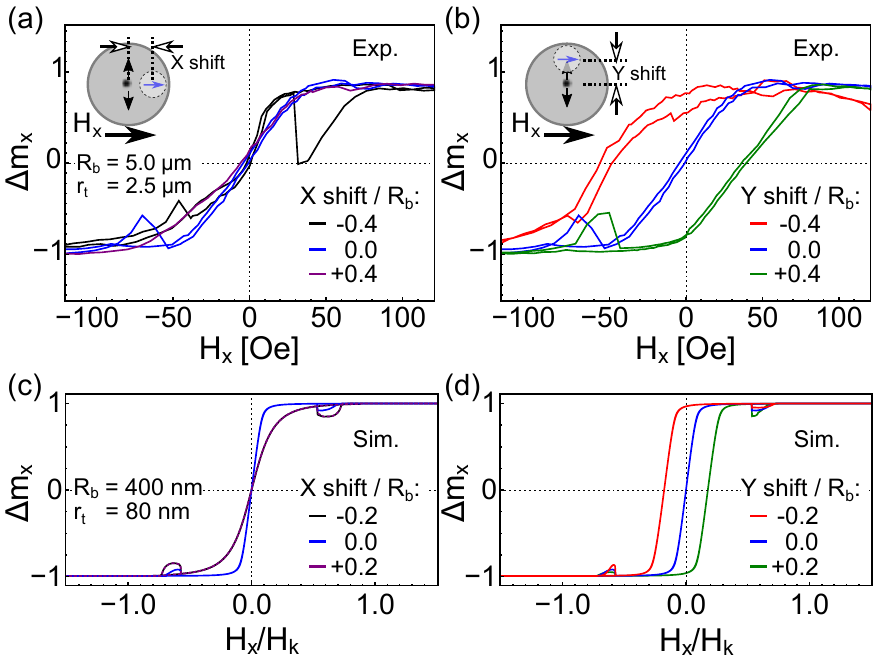}
        \caption{The effect of the shift of pinned layer disk, showing the comparison between (a, b) the experimental results, and (c, d) the simulation results; where the shifts are along: (a, c) the $X$ direction, and (b, d) the $Y$ direction.}
        \label{fig:shift}
    \end{figure}
    
    The shift of the top pinned layer disk is also important to control the sensor response. We measured and simulated the effect of shifts either parallel or transverse to $H_x$ direction [Fig.~\ref{fig:shift}]. A shift in the $X$ direction does not have a significant effect on the sensor response [Figs.~\ref{fig:shift}(a,c)]. On the other hand, the shift in $Y$ direction causes a drastic shift of the transfer curve center [Figs.~\ref{fig:shift}(b,d)]. The vortex core moves perpendicular to an applied field. Therefore, a shift of the pinned layer center ($C_Y$) in the $Y$ direction will result in a shift in the sensor transfer curve center by $H_\mathrm{shift}/H_k = C_Y/R_b$, whereas a shift in the $X$ direction does not have a significant effect [the insets in Figs.~\ref{fig:shift}(a,b)].
    We propose that a combination of four off-centered top pinned disks on a single vortex free layer would work as a multi-axis sensor. An independent readout of TMR from each top disk can be used to find the location of the vortex core, hence the direction of magnetic field in the inplane direction. Such a magnetic vector sensor would be more practical to implement, compared with the common method of using two separate sensors channels, which require mechanical alignment at orthogonal sensing axes \cite{yamazaki_2011}.

\section{Conclusions}
    We demonstrate the control and design of the sensitivity in vortex-type magnetic MTJ sensors. We used the pinned layer size geometry as an effective method to increase the sensitivity. This is another degree of freedom in design, in addition to the free layer dimensions. {We validated this approach experimentally in MTJ sensors, and with micromagnetics simulations of a single NiFe disk.} At the current demonstration, by decreasing the pinned layer diameter from {$2r_t = 9.5$ to $2 \mu$m, we could tune the effective dynamic range $2H_k$ from 156 to 31 Oe, while keeping the vortex free layer diameter fixed at $2R_b = 10 \mu$m.} {The simple linear displacement of vortex core by applied field makes the design of vortex sensors straightforward for applications; the sensor sensitivity becomes $\propto R_b^2 / r_t$. The combination of varying vortex layer and pinned layer diameters covers 2--3 orders of magnitude in sensitivity design, which makes vortex MTJ sensors as a candidate for a wide range of magnetic sensing applications.} 

\section*{Acknowledgment}
    This work was supported by the Center for Science and Innovation in Spintronics (CSIS), Center for Spintronics Research Network (CSRN), Tohoku University, the S-Innovation program, Japan Science and Technology Agency (JST), and by JSPS KAKENHI Grant Number JP19K15429.

\bibliographystyle{iopart-num}
\bibliography{refs.bib}

\newpage
\clearpage
\onecolumn
\begin{center}
    \textbf{\large Supplementary Information: \mytitle}\par
    \vspace{1em}
    {
    Motoki Endo$^{1}$, 
    Muftah Al-Mahdawi$^{2,3}$, 
    Mikihiko Oogane$^{1,2,3}$, 
    and Yasuo Ando$^{1,2,3}$
    }\\
    {
    $^{1}$Department of Applied Physics, Tohoku University, Sendai 980-8579, Japan \\
    $^{2}$Center for Science and Innovation in Spintronics (Core Research Cluster), Tohoku University, Sendai, 980-8577, Japan \\
    $^{3}$Center for Spintronics Research Network, Tohoku University, Sendai 980-8577, Japan
    }\\
    {
    E-mails: endo@mlab.apph.tohoku.ac.jp; mahdawi@tohoku.ac.jp.
    }

\end{center}\par
\vspace{1em}

\setcounter{section}{0}
\setcounter{equation}{0}
\setcounter{figure}{0}
\setcounter{table}{0}
\setcounter{page}{1}
\makeatletter
\renewcommand{\thesection}{S\arabic{section}}
\renewcommand{\theequation}{S\arabic{equation}}
\renewcommand{\thefigure}{S\arabic{figure}}
\renewcommand{\thepage}{S\arabic{page}}
\renewcommand{\@biblabel}[1]{[SR#1]}
\renewcommand\@cite[1]{[SR#1]}
\section{Magnetic properties and TMR characteristics of the sensor films}
\begin{figure}[hb]
 \begin{center}
     \includegraphics[width=0.8\textwidth]{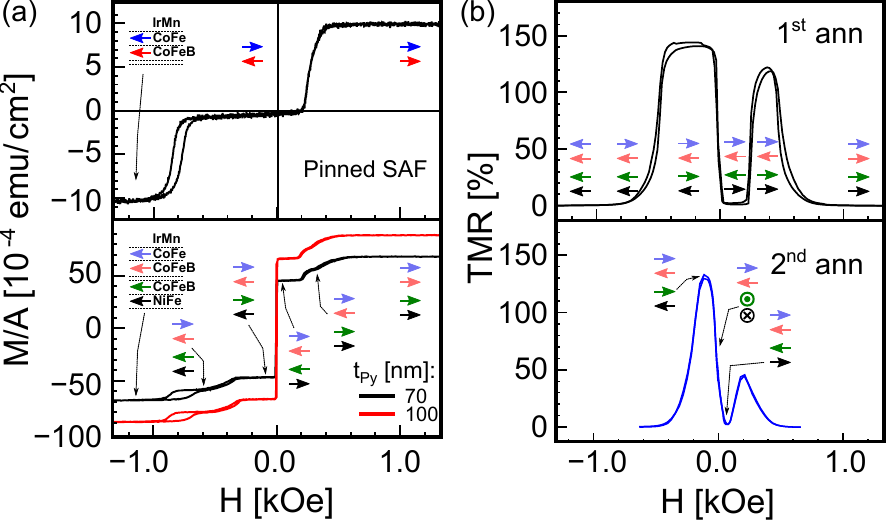}
    \caption{(a) Magnetic characteristics of the top pinned SAF (top), and MTJ films (bottom). (b) TMR characteristics after the first (top) and second (bottom) annealing steps.}
    \label{fig:film_vsm}
 \end{center}
\end{figure}

We show the magnetic properties of the blanket film in Fig.~\ref{fig:film_vsm}(a). We optimized the thickness of CoFe in the pinned SAF for zero remnant magnetization, to reduce the effect of stray fields on the vortex core in the free layer. The magnetization reversal process of the full stack is indicated by the colored arrows. Although NiFe is rather thick, the coupling in the free layer between NiFe and CoFeB is antiferromagnetic through the Ru spacer, indicating the uniformity of the Ru ultra-thin layer. Fig.~\ref{fig:film_vsm}(b) shows the TMR characteristics of large pillars after the two-step field annealing process. The MTJ pillars have a 2:1 elliptical cross section, with a major axis of 48 $\mu$m in length. After the first annealing process, we obtained a relatively high TMR ratio of 140--150 $\%$, and the transfer curve shows switching character. After the second annealing step, the pinned layer is rotated orthogonal to free layer easy axis, and a linearized transfer curve is obtained.

\newpage

\section{Induced magnetic anistropy in NiFe}

\begin{figure}[hb]
 \begin{center}
     \includegraphics[width=0.6\textwidth]{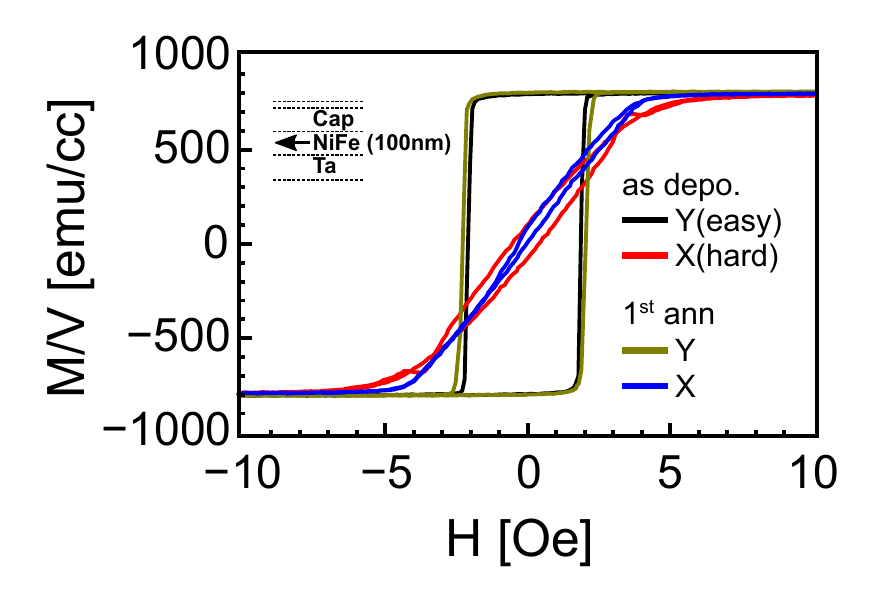}
    \caption{The magnetization loops along the easy ($Y$) and hard ($X$) axes of NiFe films in the as deposited state, and after the first annealing.}
    \label{fig:sup_aniso}
 \end{center}
\end{figure}

The NiFe exhibit an induced magnetic anisotropy due to Ni-Fe pair ordering \cite{sup_chikazumi_1955}. The pair ordering occurs during the deposition in a magnetic field \cite{sup_katada_2000}. In Fig.~\ref{fig:sup_aniso}, we show the results of magnetization loops of blanket films of: thermally-oxidized silicon substrate/Ta 5/Ru 10/Ta 5/Ni$_{80}$Fe$_{20}$ 100/MgO $1.5$/Ta 1.0, where the numbers are the nominal thicknesses in nanometers. We measured the magnetiztion loops along the easy and hard axes in the as-deposited state, and after the first pin annealing with the same conditions as the main text, \emph{i.e.}~$350^\circ\mathrm{C}$ along the easy axis of induced anisotropy in NiFe ($Y$ direction). We obtain {a saturation magnetization of 800 emu/cm$^3$, an anisotropy field of 4 Oe, and an anisotropy energy of $1.6\times 10^3$ erg/cm$^3$,} similar to the literature values \cite{sup_katada_2000}. The anisotropy field is not affected by pin annealing. However, coercivity is reduced and linearity is improved in the hard axis loop.

\newpage

\section{Critical size for vortex stability}

\begin{figure}[hb]
 \begin{center}
     \includegraphics[width=0.7\textwidth]{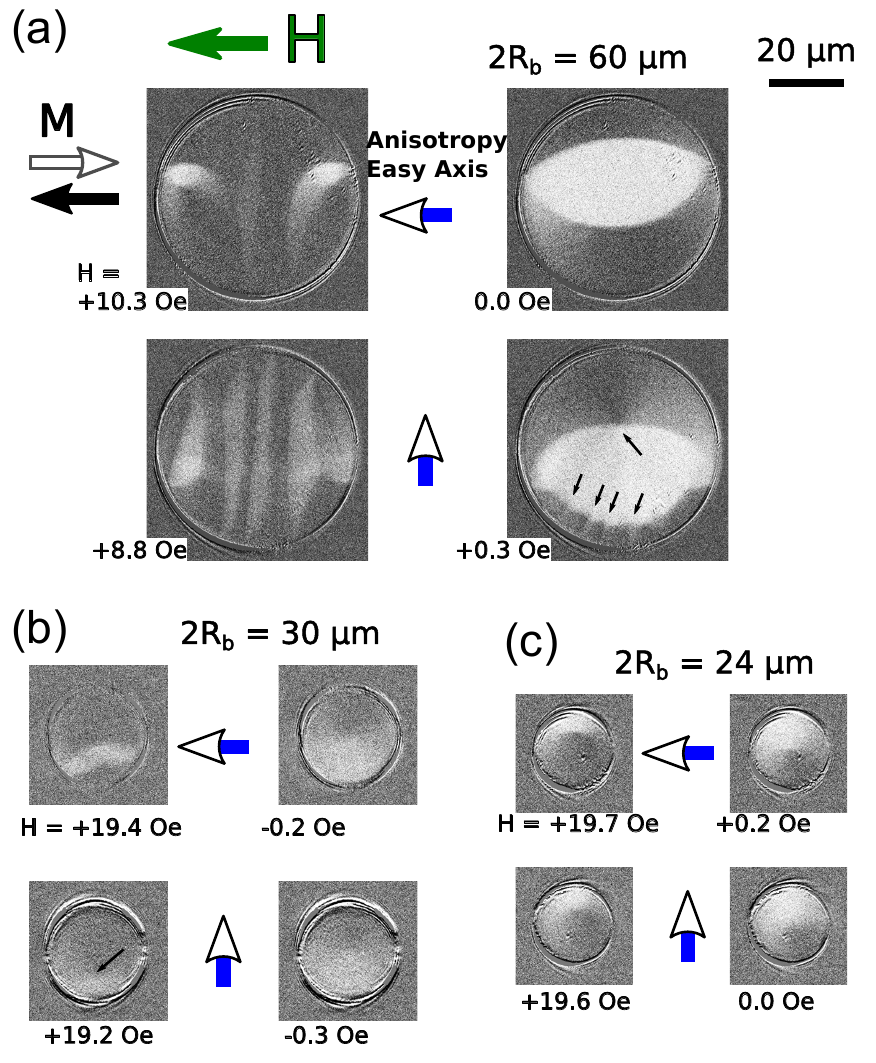}
    \caption{Longitudinal MOKE domain images of NiFe disks during field scans along the easy and hard axes directions. The disk diameters are: (a) 60 $\mu$m, (b) 30 $\mu$m, and (c) 24 $\mu$m.}
    \label{fig:sup_domain}
 \end{center}
\end{figure}

We show the effect of induced anisotropy on the domain structure in Fig.~\ref{fig:sup_domain}. We fabricated circular disks with varying diameters from the films in Sec.~S2, after the first pin annealing. We used longitudinal MOKE to image the domain structure during magnetization loops with field applied along the easy or hard axes. In large diameter disks, a reversal domain forms at zero field [Fig.~\ref{fig:sup_domain}(a)]. The initial nucleation domains form parallel to easy axis direction. When the anisotropy axis is transverse to the applied field, multiple vortices are formed [indicated by {black} arrows in Fig.~\ref{fig:sup_domain}(a)]. At a critical diameter of 30 $\mu$m, the nucleation starts from a reversal domain in the easy-axis loop, whereas the vortex state nucleates for the hard-axis loop [indicated by an arrow in Fig.~\ref{fig:sup_domain}(b)]. Below that diameter, the vortex state is the stable reversal configuration, regardless of the field direction [Fig.~\ref{fig:sup_domain}(c)].

\newpage
\section{Nucleation features in the TMR-$H$ curves}
\begin{figure}[hb]
 \begin{center}
     \includegraphics[width=0.8\textwidth]{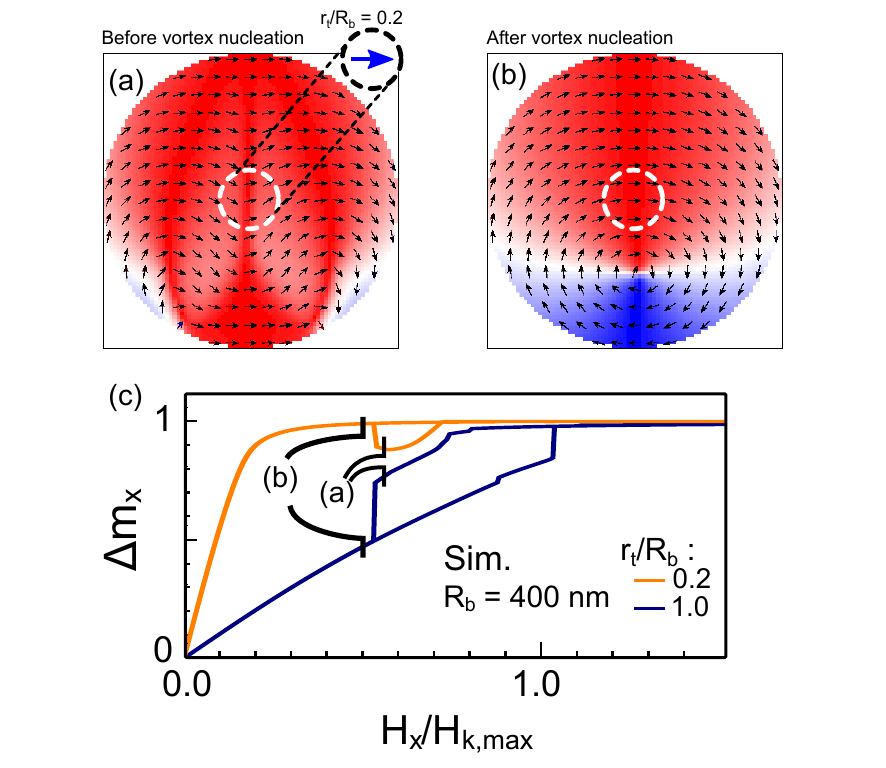}
    \caption{Simulations of the domain state before and after the vortex nucleation. (a) Before vortex nucleation, and (b) after the vortex nucleation. In (a), we show a small area to be enclosed by the pinned layer, equivalent to a projection along the $X$ direction (blue arrow). (c) The $\Delta m_x$--$H_x$ loops for a small enclosed area ($r_t/R_b = 0.2$), or a large one ($r_t/R_b = 1.0$). The states in (a) and (b) are indicated on the curves. }
    \label{fig:sup_nuc}
 \end{center}
\end{figure}

{In Figs.~3(a,d), and 4(a--d) of the main text, there are dips that appear near $\pm$50 Oe for small $r_t/R_b$ ratios. They coincide with the nucleation field of the vortex state estimated from the VSM measurements of Fig.~2(a). We show the explanation from simulation results in Fig.~\ref{fig:sup_nuc}. Below the saturation field, and before the nucleation event, a curly domain appears [Fig.~\ref{fig:sup_nuc}(a)]. The magnetization vector becomes flat at the poles of the disk to minimize the magnetostatic energy. This causes a rotation of the magnetization at the center of the disk towards a $\approx 90^\circ$ direction. After vortex nucleation, the magnetization at the center region returns back to $0^\circ$ [Fig.~\ref{fig:sup_nuc}(b)]. If the area enclosed by the pinned layer is small  [orange curve in Fig.~\ref{fig:sup_nuc}(c)], then before nucleation there is a dip in $\Delta m_x$. After nucleation, $\Delta m_x$ increases to 1, until the vortex position is close to the pinned layer edge. If the area enclosed by the pinned layer is large, then the average $\Delta m_x$ is smaller in the vortex state compared to pre-nucleation state [blue curve in Fig.~\ref{fig:sup_nuc}(c)]. }

%
%
\end{document}